\documentclass[aoas,MSNbibl,nameyear,seceqn,rotating,dvips]{arximspdf}
\usepackage{dcolumn}
\usepackage{graphicx}

\doi{10.1214/13-AOAS659} 
\volume{7}
\issue{4}
\pubyear{2013}
\firstpage{2205}
\lastpage{2228}

\makeatletter

\newcolumntype{d}[1]{D{.}{.}{#1}}

\newcommand{\rright}{\right}
\newcommand{\lleft}{\left}

\makeatother

\begin{document}
\begin{frontmatter}

\title{Principal trend analysis for time-course data with applications
in genomic medicine\thanksref{T1}}
\runtitle{Principal trend analysis}
\thankstext{T1}{Supported by P01HG000205 and NIH U54 GM-062119.}

\begin{aug}
\author[A]{\fnms{Yuping} \snm{Zhang}\corref{}\ead[label=e1]{yuping.zhang@yale.edu}}
\and
\author[B]{\fnms{Ronald} \snm{Davis}\ead[label=e2]{dbowl@stanford.edu}}
\runauthor{Y. Zhang and R. Davis}
\affiliation{Yale School of Public Health and Stanford University}
\address[A]{Department of Biostatistics\\
Yale School of Public Health\\
New Haven, Connecticut 06520-8034\\
USA\\
\printead{e1}}
\address[B]{Stanford Genome Technology Center\\
Stanford University\\
Palo Alto, California 94306\\
USA\\
\printead{e2}} 
\end{aug}

\received{\smonth{8} \syear{2011}}
\revised{\smonth{5} \syear{2013}}

%
\begin{abstract}
Time-course high-throughput gene expression data are emerging in
genomic and translational medicine. Extracting interesting time-course
patterns from a patient cohort can provide biological insights for
further clinical research and patient treatment. We propose
\emph{principal trend analysis} (PTA) to extract principal trends of
time-course gene expression data from a group of patients, and identify
genes that make dominant contributions to the principal trends. Through
simulations, we demonstrate the utility of PTA for dimension reduction,
time-course signal recovery and feature selection with high-dimensional
data. Moreover, PTA derives new insights in real biological and
clinical research. We demonstrate the usefulness of PTA by applying it
to longitudinal gene expression data of a circadian regulation system
and burn patients. These applications show that PTA can extract
interesting time-course trends with biological significance, which
helps the understanding of biological mechanisms of circadian
regulation systems as well as the recovery of burn patients. Overall,
the proposed PTA approach will benefit the genomic medicine research.
Our method is implemented into an R-package: PTA (Principal Trend
Analysis).
\end{abstract}

%
\begin{keyword}
\kwd{Time-course}
\kwd{longitudinal}
\kwd{high dimensional}
\kwd{principal trend analysis}
\kwd{sparse}
\kwd{smooth}
\kwd{principal component analysis}
\end{keyword}

\end{frontmatter}

\section{Introduction}

High-throughput technologies, such as microarray,\break  \mbox{LC-MS}, next
generation sequencing, etc., have been applied widely in current
biological and clinical studies. In the past decades, data were usually
collected at one time point. Thus, these data have three
attributes---feature (gene, protein, etc.), individual (samples,
subjects, etc.) and
value (gene expression, protein abundance, etc.), which can be
represented by matrices with each row indicating feature identity, each
column indicating sample identity and each cell recording gene
expression or protein abundance. Developing effective ways to analyze
such high-throughput genomic and proteomic data is one of the major
challenges of bioinformatics and computational biology. In such
studies, the number of features $p$ is usually much larger than the
number of samples $n$. The \emph{lasso} method \citet{Tibshirani1996}
was introduced and combined with matrix decomposition for computing a
rank-$K$ approximation for a matrix; see \citet
{zou2006sparse,mairal2010online,witten2009extensions,witten2009penalized}.\looseness=-1

The identification of time-course gene/protein expression patterns has
attracted increasing attention in biological and clinical research.
Time-course genomic and proteomic data have been collected in many
clinical and biological studies. It is common for biologists to ask the
following question on a set of genes they are interested in: what kinds
of dynamic patterns do these genes have? For example, given a group of
genes with circadian regulation functions, we want to know their
dynamic patterns and their relationships with external light signals.
This is important for understanding biological mechanisms in circadian
systems, which are critical for maintaining normal living for live
organisms. Or in a more unsupervised setting, without preselecting a
small set of genes, we want to extract underlying dominant time-course
patterns and automatically identify important genes that have
contributions on those significant patterns from a genome-wide
longitudinal data set. This kind of statistical analysis is especially
important for the research of complex diseases, which can provide a
systematic view of genomic/proteomic dynamic responses. For example, in
a study of host response to burn injuries, clinical investigators
monitored a group of burn patients and measured their gene expression
over time (\href{http://www.gluegrant.org}{www.gluegrant.org}). To
improve our systematic understanding
of the key regulatory elements in the recovery of burn patients, we
need to characterize dynamic response in gene expression. To do so, we
need a statistical learning approach for high-dimensional longitudinal
data to extract underlying time-course patterns and identify important
features that contribute to the underlying patterns of interest.

Unlike stationary gene expression data without consideration of time,
time-course gene expression data have an additional time attribute.
Traditional principal component analysis applied to stationary gene
expression data matrix for multiple samples cannot be used directly in
this scenario. Thus, dimension reduction methods for time-course data
are needed. The method that we propose draws on ideas from the
spline-based methods on time-course data analysis [\citet
{Kimeldorf1970,Wahba1990}] and principal component analysis for
dimension reduction.
Our method has the following advantages:
\begin{enumerate}
\item Unsupervised approach to automatically discover underlying gene
expression time-course patterns.
\item Automatically identifying important genes and classifying them
into different groups which contribute to different time-course patterns.
\end{enumerate}


The remainder of this article is organized as follows.
Section~\ref{section1} gives the model and algorithms for principal
trend analysis (PTA).
Section~\ref{simulation} gives the simulation studies to show the
performance of PTA at different scenarios.
Section~\ref{leaf} gives an application of PTA on a circadian
regulation gene expression data set with a subset of genes as known
targets. This data set is from arabidopsis thaliana rosette, which is a
widely-used pattern organism. Gene expression was measured six times
during 12-h-light/12-h-dark treatments of arabidopsis thaliana rosettes.
Section~\ref{burn} gives an application of PTA on a burn patient gene
expression data set without prior information of genes of interest.
This study not only shows the usefulness of PTA, but also provide
insights for the further research of burn disease.
Section~\ref{discussion} discusses the prospective of our methods and
future directions.

\section{Principal trend analysis}\label{section1}
Let $y_{\mathit{npt}}$ denote the gene $p$ expression of subject $n$ at time
point $t$, $p \in\{1,\ldots, P\}$, $n \in\{1,\ldots, N\}$, $t \in\{
1,\ldots,T\}$. For each gene $p$, $y_{p}$ was centered. We assume all the
subjects are from the same population. We want to find the
population-level time-course patterns. Thus, we propose a new method
called Principal Trend Analysis (PTA) to solve this problem.



We use the following notation. Let $\mathbf{Y}_n$ denote a $P \times T$
matrix of observations on subject $n$; $\mathbf{A}$ denote a $P\times
K$ matrix of factor scores, $\mathbf{A} = [\mathbf{a}_{1},\ldots,
\mathbf{a}_{K}]$, $[\mathbf{A}]_{p,k} = a_{p,k}$; $\bolds{\Theta}$
denote a $K \times(T+2)$ matrix of spline coefficients, $\bolds{\Theta
} = [\bolds{\theta}_{1}^{\mathsf{T}},\ldots, \bolds{\theta
}_{K}^{\mathsf{T}}]$, $[\bolds{\Theta}]_{k,m}=\theta_{k,m}$; $\mathbf
{B}$ denote a $T \times(T+2)$ matrix containing the cubic spline
basis, $[\mathbf{B}]_{t,m}=B_{m}(t)$; $\mathbf{S}$ denote a $K \times
T$ matrix presenting top $K$ time-course trends across $T$ time points,
$[\mathbf{S}]_{k,t}= S_{k}(t)$; and $\mathbf{S}=\bolds{\Theta}\mathbf
{B}^\mathsf{T}$.

The underlying model is as below:
%
%
\begin{equation}
\label{modelmatrix} \mathbf{Y} = \lleft[ \matrix{
\mathbf{Y}_{1}
\cr
\mathbf{Y}_{2}
\cr
\vdots
\cr
\mathbf{Y}_{N}} \rright] =\lleft[ \matrix{\mathbf{A}\bolds{\Theta}
\mathbf{B}^\mathsf{T}
\cr
\mathbf{A}\bolds{\Theta}\mathbf{B}^\mathsf{T}
\cr
\vdots
\cr
\mathbf{A}\bolds{\Theta}\mathbf{ B}^\mathsf{T}} \rright]+\lleft[
\matrix{\mathbf{E}_{1}
\cr
\mathbf{E}_{2}
\cr
\vdots
\cr
\mathbf{E}_{N}} \rright].
\end{equation}

Let $\hat{\bolds{\Theta}}$ and $\hat{\mathbf{A}}$ denote the parameter
estimates of $\bolds{\Theta}$ and $\mathbf{A}$ for model
(\ref{modelmatrix}). Since both $\bolds{\Theta}$ and $\mathbf{A}$ are
unknown, we cannot obtain their least squares solutions simultaneously.
Thus, we propose an iterative algorithm to estimate the parameters by
solving the following optimization problem:
%
%
\begin{eqnarray}
\label{matrixop}
&&\min_{\mathbf{A}, \bolds{\Theta}} L(\mathbf{A},
\bolds{\Theta}|\mathbf{Y})\quad\mbox{subject to}\nonumber\\[-8pt]\\[-8pt]
&&\bolds{\Theta}\bolds{\Omega}
\bolds{\Theta}^\mathsf{T} \le c_1,\qquad \|\mathbf{A}
\|_1\le c_2 \quad\mbox{and}\quad\|\mathbf{A}
\|_2^2 = 1,\nonumber
\end{eqnarray}
where $L(\mathbf{A}, \bolds{\Theta}|\mathbf{Y}) = \sum_{n=1}^{N}\|
\mathbf{Y_n} - \mathbf{A}\bolds{\Theta}\mathbf{ B}^\mathsf{T} \|_F^2$
is a loss function, $\|\cdot\|_F$ is the Frobenius norm, $\|\mathbf{A}\|
_1$ is the nuclear norm of $\mathbf{A}$, $\|\mathbf{A}\|_2$ is the
euclidean norm of $\mathbf{A}$, $\bolds{\Omega}$ denotes a $(T+2)
\times(T+2)$ matrix, and
\[
\Omega_{ij} = \int B_i^{\prime\prime}(t)B_{j}^{\prime\prime}(t)
\,dt.
\]

Small values of $c_1$ produce smoother curves while larger values
produce more wiggly curves. At the one extreme, as $c_1 \rightarrow0
$, the penalty term dominates, forcing $S_{k}^{\prime\prime}(t)=0$ everywhere,
and thus the solution is the least-square line. At the other extreme,
as $c_1 \rightarrow\infty$, the penalty term becomes unimportant and
the solution tends to be an interpolating twice-differentiable
function.\vspace*{1pt}

When $1\le c_2 \le\sqrt{P}$, we obtain sparsity on genes. When $c_2 >
\sqrt{P}$, the Lasso-penalty will be inactive in the condition that $\|
\mathbf{A}\|_2^2 \le1$.

The optimization problem (\ref{matrixop}) is not convex due to the
$L_2$-equality penalty on~$\mathbf{A}$. We modify the $L_2$-equality
penalty in (\ref{matrixop}) and obtain the following optimization problem:
%
%
\begin{eqnarray}
\label{matrixopmodify}
&&
\min_{\mathbf{A}, \bolds{\Theta}}
L(\mathbf{A}, \bolds{\Theta}|\mathbf{Y})\quad\mbox{subject
to}\nonumber\\[-8pt]\\[-8pt]
&&\bolds
{\Theta}
\bolds{\Omega} \bolds{\Theta}^\mathsf{T} \le c_1,\qquad \|
\mathbf{A}\|_1\le c_2 \quad\mbox{and}\quad\|\mathbf{A}
\|_2^2 \le1.\nonumber
\end{eqnarray}

Using Lagrange multipliers, we rewrite the criteria in (\ref{matrixop})
and (\ref{matrixopmodify}) as
%
%
\begin{equation}
\label{dual} \min_{\mathbf{A}, \bolds{\Theta}} L(\mathbf{A}, \bolds
{\Theta}|\mathbf
{Y})+\lambda_1 \bolds{\Theta} \bolds{\Omega} \bolds{
\Theta}^\mathsf{T} +\lambda_2 \|\mathbf{A}\|_1 +
\lambda_3 \|\mathbf{A}\|_{2}^{2}.
\end{equation}

The supplement S1 [\citet{ZhaDav13}] proves that the optimization problem (\ref
{matrixopmodify}) is biconvex, so it can be solved with an iterative
algorithm. Moreover, the solution to (\ref{matrixopmodify}) also
satisfies $\|A\|_2^2=1$, provided that $\lambda_3$ is chosen so that
(for fixed $\bolds{\Theta}$) the $\mathbf{A}$ that maximizes $\min
_{\mathbf{A}} L(\mathbf{A}|\bolds{\Theta},\mathbf{Y})$,
subject to $\|\mathbf{A}\|_1\le c_2$,
has $L_2$-norm greater than or equal to 1. This follows from the
Karush--Kuhn--Tucker conditions in convex optimization [\citet
{boyd2004convex}]. Thus, for appropriately chosen $\lambda_3$, the
solutions to (\ref{matrixopmodify}) solve (\ref{matrixop}).


If preselected genes are available (e.g., we are interested in one
particular pathway or genes with the same biological functions), and
the problem is to extract the underlying key time-course trends for
these genes, we can remove the \emph{lasso} penalty in the optimization
problem (\ref{dual}), that is, let $\lambda_2=0$.



\subsection{Algorithm}


When $K=1$, we compute the rank-$1$ sparse principal trend (PT) as
follows (cf. S2 in the supplementary material [\citet{ZhaDav13}] for the derivation):
\begin{enumerate}
\item Initialize $\bolds{\theta}^{(1)}$ and $\mathbf{a}^{(1)}$ with $\|
\mathbf{a}\|_{2}=1$.
\item For $i=1,2,\ldots\,$, until convergence:
\begin{enumerate}[(a)]
\item[(a)] $\bolds{\theta}^{(i+1)} \leftarrow(N \mathbf{B}^{\mathsf
{T}}\mathbf{B}\otimes\mathbf{a}^{(i)\mathsf{T}}\mathbf{a}^{(i)}+
\lambda_1 \bolds{\Omega})^{-1} \mathbf{B}^{\mathsf{T}}
( \sum_{n=1}^{N} \mathbf{Y}_n )\mathbf{a}^{(i)} $.
\item[(b)]
When $\lambda_2 =0$,\\[2pt]
$\mathbf{a}^{(i+1)} \leftarrow( \sum_{n=1}^{N}
\mathbf{Y}_n ) \mathbf{B} \bolds{\theta}^{(i)} $,\\[2pt]
$\mathbf{a}^{(i+1)} \leftarrow\frac{ \mathbf{a}^{(i+1)}}{\|\mathbf
{a}^{(i+1)}\|_{2}}$.\\[2pt]
When $\lambda_2>0$,\\
$\mathbf{a}^{(i+1)} \leftarrow\operatorname{Soft}( (\sum_{n=1}^{N}
\mathbf{Y}_n ) \mathbf{B} \bolds{\theta}^{(i)}, \frac{1}{2}\lambda
_2 )$, where $\operatorname{Soft}(\cdot)$ denote the soft thresholding
operator; that is, $\operatorname{Soft}(x,c)=
\operatorname{sgn}(x)(|x|-c)_{+}$, where $c >0$ is a constant and
$x_{+}$ is defined to equal $x$ if $x>0$ and 0 if $x\le0$.\\
$\mathbf{a}^{(i+1)} \leftarrow\frac{ \mathbf{a}^{(i+1)}}{\|\mathbf
{a}^{(i+1)}\|_{2}}$.
\end{enumerate}
\end{enumerate}

For multiple component decomposition with $1<K\le\min\{P, T\}$, we
compute the rank-$K$ sparse time-course PT analogously.

For $k \in\{1,\ldots, K\}$, iterate the following procedure:
\begin{enumerate}
\item Let $\mathbf{Y}_n^{1} \leftarrow\mathbf{Y}_n$, for every $n \in
\{1,\ldots, N\}$.
\item For $k \in\{1,\ldots, K\}$:
\begin{enumerate}[(a)]
\item[(a)] Obtain\vspace*{1pt} $\bolds{\theta}_{k}$ and $\mathbf{a}_{k}$ using the
single component decomposition algorithm.
\item[(b)] $\mathbf{Y}_n^{k+1} \leftarrow\mathbf{Y}_n^{k} - \mathbf
{a}_{k}\bolds{\theta}_{k}\mathbf{B}^\mathsf{T}$.
\end{enumerate}
\end{enumerate}

\begin{figure}

\includegraphics{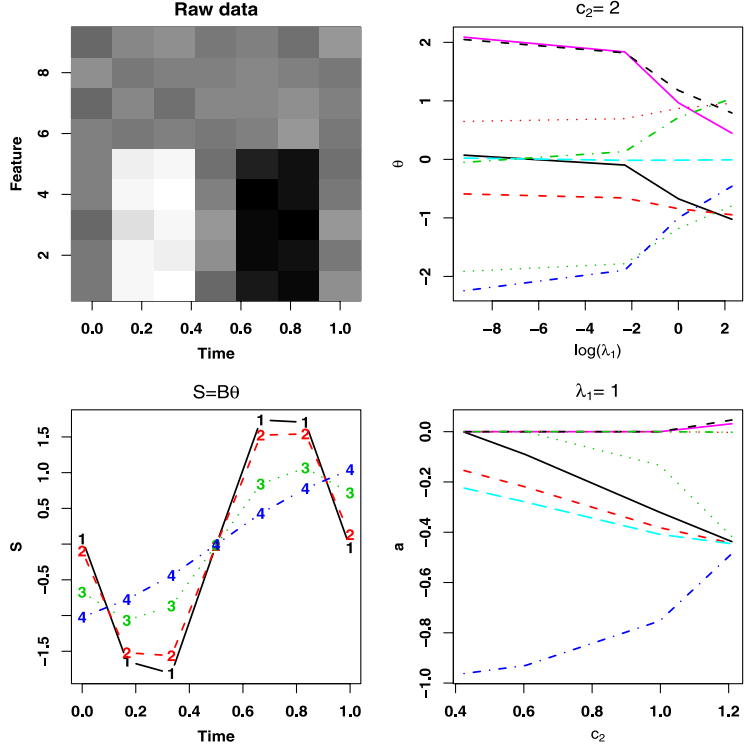}

\caption{Effects of tuning parameters for PTA. Top-left panel: raw
data, with each row indicating one feature, each column indicating one
time point; top-right panel: paths of the coefficient vector
$\bolds{\theta}$ according to changes of tuning parameter $\lambda_1
(\lambda_1 \in\{0.0001, 0.1, 1, 10\})$ with $\lambda_2$ fixed to 2;
bottom-left panel: four estimated time-course trends ($\mathbf{S} =
\mathbf{B}\bolds{\theta}$, indicated by 1, 2, 3, 4) according to the
coefficients and parameters in top-right panel with $\lambda_1$
equaling to 0.0001, 0.1, 1 and 10, respectively; bottom-right panel:
paths of coefficient vector $a$ according to changes of tuning
parameter $c_2$ with $\lambda_1=1$.}
\label{figure0}\vspace*{6pt}
\end{figure}

The PTA algorithm is not guaranteed to get the global minimum similar
to \citet{de1994block}, but behaves well in practice. We
illustrate the
effect of parameters by a simple example. We simulate a data set with 9
features, 7 time points and 1 replicate. Values for the first 5
features are drawn from $\sin(2\pi t) + N(0, 0.1)$, and those for the
remaining 4 features are drawn from $N(0, 0.1)$. Raw data is shown in
the top-left panel of Figure~\ref{figure0}, with each row indicating
one feature and each column indicating one time point. The top-right
%
%
panel of Figure~\ref{figure0} shows how the coefficient vector $\bolds
{\theta}$ changes according to tuning parameter $\lambda_1 (\lambda_1
\in\{0.0001, 0.1, 1, 10\})$ with $\lambda_2$ fixed at 2. The
bottom-left panel of Figure~\ref{figure0} shows the estimated
time-course trends with $\lambda_1$ equal to 0.0001, 0.1, 1~and~10,
respectively. One can see that smaller values of $\lambda_1$ (larger
values of~$c_1$) produce more wiggly curves, while larger values of
$\lambda_1$ (smaller values of $c_1$) produce smoother curves. The
bottom-right panel of Figure~\ref{figure0} shows how the coefficient
vector $a$ changes according to the tuning parameter $c_2$ with $\lambda
_1$ fixed to~1. One can see that smaller values of $c_2$ (larger values
of $\lambda_2$) produce more sparsity of~$a$.\looseness=1

\subsection{PTA for missing data}
The PTA works in the case of missing data. When some elements of the
data $y_{\mathit{npt}}$ ($n \in\{1,\ldots,N\}$, $p \in\{1,\ldots,P\}$, $t \in\{
1,\ldots,T\}$) are missing, those elements can simply be excluded from all
computations. Let $C$ denote the set of indices of nonmissing elements
in $\mathbf{Y}$. The criterion is as follows:
%
%
\begin{equation}
\label{globalone} \arg\min_{\mathbf{A}, \bolds{\Theta}}\Biggl\{\sum
_{(n,p,t)\in C} \Biggl[y_{\mathit{npt}} - \sum
_{k=1}^{K}a_{pk}S_{k}(t)
\Biggr]^{2} \Biggr\},
\end{equation}
subject to $\bolds{\Theta}\bolds{\Omega}\bolds{\Theta}^\mathsf{T}
\le c_1$, $\|\mathbf{A}\|_2^2 \le1$, and $\|\mathbf{A}\|_1 \le c_2$,
where $\mathbf{A}$ is the matrix consisting of elements $a_{pk}$ and
$\bolds{\Theta}$ is the matrix consisting of elements $\theta_{km}$.
When the observed samples cover all the time points and genes of
interest, this approach will work well. Admittedly, if there are too
many missing observations (e.g., there is no observed data for one or a
few genes), it may cause problems.

\subsection{Tuning parameter selection for PTA}
In PTA, the tuning parameters are $c_1$ in the constraint $\bolds
{\Theta}\bolds{\Omega}\bolds{\Theta}^\mathsf{T} \le c_1$ for
smoothing, and $c_2$ in the constraint $\|\mathbf{A}\|_1\le c_2$ for
feature selection. We use cross-validation to select appropriate tuning
parameters, as cross-validation (CV) is a simple and widely used method
for estimating prediction error [cf. \citet{Hastie2009} for a
description of CV and selection of tuning parameters]. The algorithm of
PTA to select tuning parameters is as follows:

\begin{itemize}
\item From the original data matrix $\mathbf{Y}$, construct $m$ data
matrices, where $m$ is the number of folds stratified for
cross-validation. Let $\mathbf{Y}^{1},\ldots, \mathbf{Y}^{m}$ be subsets
of $\mathbf{Y}$, each of which extracts a nonoverlapping $\frac{1}{m}$
of the elements of $\mathbf{Y}$. The extracted elements are sampled at
random from entries in $\mathbf{Y}$. We treat those extracted elements
as ``missing.''
\item For each pair of candidate values of $c_1$ and $c_2$ ($1 < c_2 <
\sqrt{P}$):
\begin{enumerate}
\item For $j \in1,\ldots, m$:
\begin{enumerate}[(a)]
\item[(a)] Fit the PTA to $\mathbf{Y}^{j}$ with the tuning parameter $c$
and calculate $\hat{\mathbf{Y}}^{j}$, the resulting estimate of $\mathbf
{Y}^{j}$. Each sample $i$ of $\mathbf{Y}^{j}$ is estimated by $\bolds
{\hat{A}}\bolds{\hat{\Theta}}\mathbf{B}^\mathsf{T}$.
\item[(b)] Record the mean squared error of $\hat{\mathbf{Y}}^{j}$. This
mean squared error is obtained by computing the mean of the squared
differences between elements of $\mathbf{Y}^{j}$ and the corresponding
elements of $\hat{\mathbf{Y}}^{j}$.
\end{enumerate}
\item Record the average mean squared error across $\mathbf
{Y}^{1},\ldots,
\mathbf{Y}^{m}$ for the tuning parameters $c_1$ and $c_2$.
\end{enumerate}
\item The optimal values of $c_1$ and $c_2$ are those that correspond
to the lowest mean squared error (with one standard error rule).
\end{itemize}

For the number of folds stratified for cross-validation, one should
choose the appropriate value which depends on the application. With
leave-one-out cross-validation, the cross-validation estimator is
approximately unbiased for the true prediction error, but variance can
be high, and the computational burden is also considerable. With five-
or ten-fold, say, cross-validation has lower variance but more bias,
depending on how the performance of the learning method varies with the
size of the training set. If the learning curve has a large slope at
the given training set size, five- or ten-fold cross-validation will
overestimate the true prediction error. Whether this bias is a drawback
in practice depends on the objective. Overall, five- or ten-fold
cross-validation is recommended as a good compromise [see \citet
{breiman1992submodel,kohavi1995study}].

\subsection{Connection with principal component analysis (PCA)}

Assume that we have $N$ $P$-dimensional data vectors $x_1, x_2,\ldots,
x_N$, which form the $P \times N$ data matrix $\mathbb{X}=[x_1,\ldots,
x_N]$. The matrix $X$ is decomposed into
%
%
\begin{equation}\label{pca}
\mathbb{X} = \mathbb{A} \mathbb{S} + \mathbf{E},
\end{equation}
where $\mathbb{A}$ is a $P \times K$ matrix, $\mathbb{S}$ is a $K
\times N$ matrix and $K \le\min{(P, N)}$, and $\mathbf{E}$ is a $P
\times N$ matrix representing the error term. Principal subspace
methods find $\mathbb{A}$ and $\mathbb{S}$ such that the reconstruction error
%
%
\begin{equation}
\|\mathbb{X}-\mathbb{A}\mathbb{S}\|_F^2 = \sum
_{p=1}^{P}\sum_{n=1}^{N}
\Biggl(x_{pn}-\sum_{k=1}^{K}a_{pk}s_{kn}
\Biggr)^2
\end{equation}
is minimized. There $F$ denotes the Frobenius norm, and $x_{pn}$,
$a_{pk}$ and $s_{kn}$ denote elements of the matrices $\mathbb{X}$,
$\mathbb{A}$ and $\mathbb{S}$, respectively. The subspace spanned by
the column vectors of the matrix $\mathbb{A}$ is called the principal
subspace. Values in each column of $\mathbb{A}$ are called scores for
that principal component. 

In the PTA model (\ref{modelmatrix}), we also use the Frobenius norm as
the loss function. $\mathbf{A}$ in mode (\ref{modelmatrix}) is also a $P
\times K$ matrix and its column vectors span the principal subspace.
However, $\mathbf{S}$ is a matrix of $K \times T$ to characterize the
data properties over time. We borrow the name for $\mathbb{A}$ used in
PCA model (\ref{pca}), and name the values in each column of $\mathbf{A}$
as scores. 

\subsection{Connections with elastic net}
Suppose that the data set has $N$ observations with $P$ predictors. Let
$\mathbb{y} = (y_1,\ldots, y_n)^\mathsf{T}$ be the response and
$\mathbb{X}$ be the $N \times P$ model matrix. After centering the
response and standardizing the predictors, \citet
{zou2005regularization} propose the elastic net to solve the following
optimization problem:
%
%
\begin{equation}
\min_{\beta}\bigl\{L( \beta)\bigr\} = \min_{\beta}
\bigl\{\|\mathbb{y}-\mathbb{X}\mathbb{\beta}\|_F^2
\bigr\},
\end{equation}
subject to $ \|\mathbb{\beta}\|_2^2 \le c_1$ and $\|\mathbb{\beta}\|_1
\le c_2$, by penalizing the coefficient vector $\beta$ using a
combination of $L_1$- and $L_2$-norm constraints. In the PTA
optimization problem~(\ref{matrixopmodify}), the penalty on $\mathbf
{A}$, which is a combination of $\|\mathbf{A}\|_1 \le c_2$ and $\|
\mathbf{A}\|_2^2 \le1$, is an elastic net type penalty. To make\vspace*{1pt} both
$L_1$- and $L_2$-constraints to be active, $c_2$ must be between 1 and
$\sqrt{P}$.

\section{Simulation study}\label{simulation}

To illustrate the performance of our method, we design the following
experiment. We simulate a data set $\mathbf{Y}$ with $P$ genes, $N$
subjects and $T$ time points. We assign the value of $y_{\mathit{npt}}$ as follows:
%
%
\begin{eqnarray}\label{simequ1}
y_{\mathit{npt}} &=& w_{0,p}\sin(2.0\cdot\pi\cdot t ) +w_{1,p}
\sin(1.0 \cdot\pi\cdot t )\nonumber\\[-8pt]\\[-8pt]
&&{} + w_{2,p}\sin(0.5\cdot\pi\cdot t )+
\epsilon_{\mathit{npt}},\nonumber
\end{eqnarray}
where $p$ is the indicator of gene, $n$ is the indicator of subject,
$t$ is time, $\epsilon_{\mathit{npt}}$~is the error term, $w_{0,p}$ is $I(0<p\le
150)$, $w_{1,p}$ is $I(150<p\le250)$, $w_{2,p}$ is $I(250<p\le300)$,
and $I$ denotes the indicator function. For each study, we repeat the
simulation 10 times and report the averages of performances and their
standard deviations.

First, we study how the performance of the proposed method changes when
the percentage of noisy features increases. Noisy features widely exist
in real applications. For instance, the human genome contains over
20,000 genes, not all of which are expressed at the same time. Even
after pre-filtering by variance or the coefficient of variance, noisy
features still exist. This is shown in the burn patient data set that
we consider in Section~\ref{burn}. Thus, we design simulations in the
presence of noisy features as illustrated in Table~\ref{design1}. Let
%
%
\begin{table}
\tabcolsep=0pt
\caption{Design 1:
studying how the performance of PTA changes when the percentage of
noisy features increases. Simulations(\protect\ref{simequ1})} are performed according to

\label{design1}
%
{\fontsize{8.5pt}{11pt}\selectfont{
\begin{tabular*}{\tablewidth}{@{\extracolsep{\fill}}l cccccccc@{}} 
\hline
\textbf{Case} & $\bolds{P}$ & $\bolds{\mathcal{P}}$ & $\bolds{N}$
& $\bolds{T}$ & $\bolds{\mathcal{P}/{P}}$ & $\bolds{\epsilon
_{\mathit{npt}}}$ & \textbf{Nonzero features} & \textbf{Explained variance}\\
\hline
1 & \hphantom{0}400 & 100 & 1 & 50 & 0.25 & $N(0, 0.1)$ & (150.0, 101.4, 50.4) $\pm
$ (0.00, 1.51, 0.97) &0.838 $\pm$ 0.0005\\ 
2 & \hphantom{0}500 & 200 &1 & 50 & 0.40 & $N(0, 0.1)$ & (150.1, 100.0, 50.0) $\pm$
(0.32, 0.00, 0.00) &0.833 $\pm$ 0.0005\\
3 & \hphantom{0}600 & 300 & 1 & 50 & 0.50 & $N(0, 0.1)$ & (150.5, 100.7, 50.0) $\pm
$ (1.58, 1.25, 0.00)&0.827 $\pm$ 0.0002\\
4 & 1000 & 700 & 1 & 50 & 0.70 & $N(0, 0.1)$ & (150.4, 100.6, 50.1) $\pm
$ (0.84, 0.52, 0.32) & 0.807 $\pm$ 0.0004\\
\hline
\end{tabular*}}}
%
\end{table}
$\mathcal{P}$ denote the number of noisy features. The percentage of
noisy features $\mathcal{P}/P$ changes from 0.25 to 0.70. We fix the
number of subjects and the number of time points, and assume the error
term follow a normal distribution $N(0, 0.1)$. We independently run the
simulation 10 times and calculate the mean and standard deviation of
the percentage of explained variance. One can see that the smaller
$\mathcal{P}/P$ is, the larger the percentage of explained variance is.
We use the proposed cross-validation method to select the number of
nonzero features for each PT at each run of the simulation. The results
are shown in the column of ``Nonzero features''. The first pair of
braces illustrates the average of nonzero features for each PT in each
case. The second pair shows the standard errors of the nonzero
features. We also calculate the percentage of explained variance. The
average of the percentage of explained variances across 10 simulations
is shown in the ``Explained variance'' column. One can see that with up
to 70\% of noisy features, our method has good performance on selecting
the true number of nonzero features that carry time-course signals.

%
%
\begin{table}
\tabcolsep=0pt
\caption{Design 2: studying how the
performance of PTA changes when the number of subjects increases.
Simulations are performed according to (\protect\ref{simequ1})}
\label{design2}
{\fontsize{8.5pt}{11pt}\selectfont{
\begin{tabular*}{\tablewidth}{@{\extracolsep{\fill}}l ccccccc@{}} 
\hline
\textbf{Case} & $\bolds{P}$ & $\bolds{\mathcal{P}}$
& $\bolds{N}$ &$\bolds{T}$& $\bolds{\epsilon_{\mathit{npt}}}$ &
\textbf{Nonzero features} &\textbf{Explained variance}\\
\hline
1 & 400 & 100 & 1 & 50 & $N(0, 0.1)$ & (150.0, 101.4, 50.4) $\pm$
(0.00, 1.51, 0.97) &0.838 $\pm$ 0.0005 \\ 
2 & 400 & 100& 5 & 50 & $N(0, 0.1)$ & (150.0, 100.0, 50.0) $\pm$ (0.00,
0.00, 0.00) &0.940 $\pm$ 0.0001 \\
3 & 400 & 100 & 10 & 50 & $N(0, 0.1)$ & (150.0, 100.0, 50.0) $\pm$
(0.00, 0.00, 0.00)&0.956 $\pm$ 0.0001\\
4 & 400 & 100 & 40 & 50 & $N(0, 0.1)$ &(150.0, 100.0, 50.0) $\pm$
(0.00, 0.00, 0.00) & 0.969 $\pm$ 0.0000\\
\hline
\end{tabular*}}}
%
\end{table}

\begin{figure}

\includegraphics{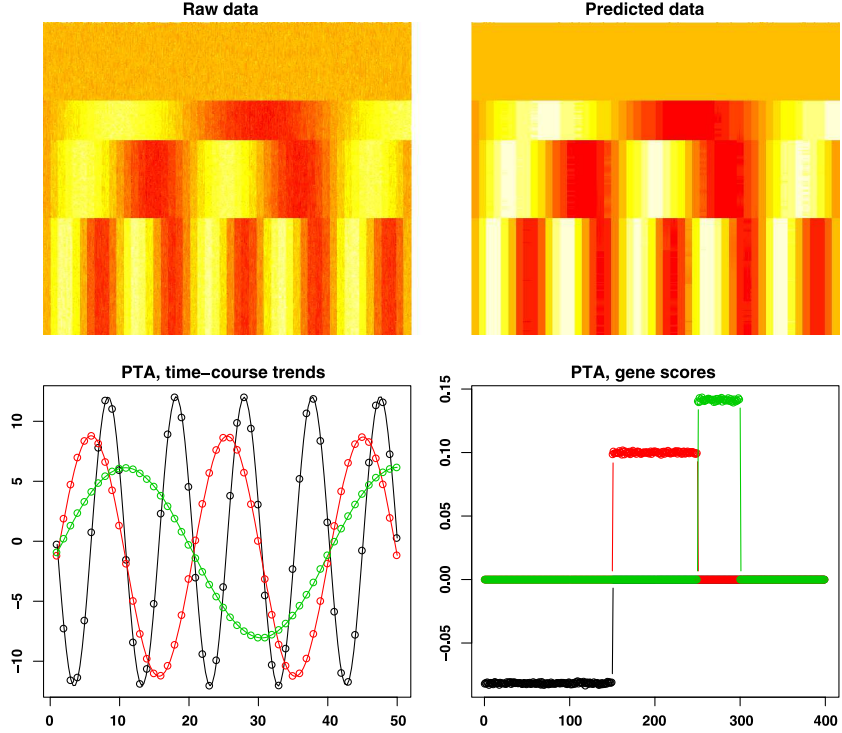}

\caption{One example in design 2. PTA on the simulated data with
multiple patterns and noisy genes. Data was simulated according to
(\protect\ref{simequ1}) with $\epsilon_{\mathit{npt}} \sim N(0, 0.1)$, $P=400$,
$N=10$ and $T=50$. The top-left panel shows the simulated raw data,
with rows indicating genes, columns indicating samples ordered by time
points and samples from the same time point grouped together. The
top-right panel shows the predicted data with three time-course trends.
The bottom-left panel shows top three PTs of PTA with three types of
frequencies: black, the first PT which extracts the dominant frequency
$f_{1}=1$; red, the second PT which extracts the second dominant
frequency $f_{2}=0.5$; green, the third PT which extracts the third
dominant frequency $f_{3}=0.25$. The bottom-right panel shows scores of
PTA on the simulated time-course gene expression data with three types
of frequencies: black, the first PT; red, the second PT; green, the
third PT. Tuning parameters are obtained by tenfold cross-validation.}
\label{figure1}
\end{figure}

Second, we study how the performance of the proposed method changes
when the number of subjects increases. In real applications, the number
of subjects varies a lot. For instance, the circadian rhythm data set
that we consider in Section~\ref{leaf} has 3 subjects, while the burn
patient data set we consider in Section~\ref{burn} has 28 subjects.
Thus, we want to investigate the performance of PTA when the number of
subjects changes. We run the simulations with different numbers of
subjects. The design of the simulation is shown in Table~\ref{design2}.
We change the number of subjects from 1 to 40 with the
number of features, the number of noisy features and the number of time
points fixed. The error term is drawn from $N(0, 0.1)$. For each case,
we repeat the simulation 10 times. One can see that as the number of
subjects increases, the performance of our method improves on both the
percentage of explained variance and the accuracy of detecting
informative features. We plot one example as shown in Figure~\ref{figure1}. The top-left heatmap shows the raw data. The top-right
heatmap shows the prediction of PTA. The bottom 250 genes of the
heatmap reflect three time-course patterns with noise filtered. The
remaining genes of the heatmap have zero values because noise has been
filtered. The bottom-left panel of Figure~\ref{figure1} shows PTs
identified by PTA. We can see that the first PT successfully extracts
the dominant frequency belonging to the first 200 genes; the second PT
extracts the second dominant frequency belonging to the second 100
genes; the third PT extracts the third dominant frequency belonging to
the remaining 50 genes. We also plot the scores in the right-bottom
panel of Figure~\ref{figure1}, which reflect the contributions of each
gene on the time-course patterns. In the first PT, the first 150
features have nonzero scores, while the remaining features have zero
scores. In the second PT, the features from 151 to 250 have nonzero
features, but the remaining features have zero scores. In the third PT,
the features from 251 to 300 have nonzero features, while the remaining
features have zero scores. The results demonstrate that our PTA method
works well on the data set with multiple time-course patterns and noisy
genes. Under the designed model (\ref{simequ1}) and parameters
in Table~\ref{design2}, PTA shows good performance even with a small
sample size.

%
%

Third, we study the relationship between the performance of PTA and the
signal-to-noise ratio. Genomic technologies such as Microarray and
RNA-seq have measurement errors. It affects the detection of true
signals in real data sets. It is necessary to investigate the
performance of PTA with different levels of signal-to-noise ratio. We
fix the number of features, the number of noninformative features, the
number of subjects and the number of time points, while increasing the
noise variance in (\ref{simequ1}) from 0.1 to 4 as shown in Table~\ref{design3}. Based on the simulation
%
%
\begin{table}
\tabcolsep=0pt
\caption{Design 3: studying the relationship between the performance of
PTA and the signal-to-noise ratio. Simulations are performed according
to (\protect\ref{simequ1})} 
\label{design3} 
{\fontsize{8.5pt}{11pt}\selectfont{
\begin{tabular*}{\tablewidth}{@{\extracolsep{\fill}}l cccccccc@{}} 
\hline
\textbf{Case} & $\bolds{P}$ & $\bolds{\mathcal{P}}$ & $\bolds{N}$ &$\bolds{T}$
& $\bolds{\epsilon_{\mathit{npt}}}$ & \textbf{SNR} & \textbf{Nonzero
features} & \textbf{Explained variance}\\ 
\hline
1 & 400 & 300 & 1 & 50 & $N(0, 0.1)$ & 7.07 & (150.0, 101.4, 50.4) $\pm
$ (0.00, 1.51, 0.97) & 0.838 $\pm$ 0.0005\\ 
2 & 400 & 300 & 1 & 50 & $N(0, 0.2)$ & 3.54 & (150.6, 100.8, 50.5) $\pm
$ (0.84, 1.14, 0.71) & 0.778 $\pm$ 0.0007\\ 
3 & 400 & 300 & 1 & 50 & $N(0, 0.35)$ & 2.02 & (151.2, 101.0, 50.3) $\pm
$ (2.04, 1.41, 0.95) & 0.649 $\pm$ 0.0021\\
4 & 400 & 300 & 1 & 50 & $N(0, 0.7)$ & 1.01 & (149.7, 100.0, 49.9) $\pm
$ (1.42, 2.62, 3.60) & 0.374 $\pm$ 0.0030\\
5 & 400 & 300 & 1 & 50 & $N(0, 1)$ & 0.71 & (147.7, 98.0, 47.7) $\pm$
(3.97, 3.33, 3.06) & 0.230 $\pm$ 0.0057\\
6 & 400 & 300 & 1 & 50 & $N(0, 2)$ & 0.35 & (145.9, 95.5, 45.0) $\pm$
(2.23, 1.58, 0.00) & 0.081 $\pm$ 0.0031\\
7 & 400 & 300 & 1 & 50 & $N(0, 4)$ & 0.18 & (145.0, 95.0, 45.0) $\pm$
(0.00, 0.00, 0.00) & 0.047 $\pm$ 0.0013 \\
\hline
\end{tabular*}}}
\end{table}
(\ref{simequ1}), we calculate the signal-to-noise ratio $\mathrm{SNR}= \frac
{1}{\sqrt{2}\sigma_{\epsilon}}$. One can see that the larger the
signal-to-noise ratio is, the more accurate the detected informative
features are. Also, the larger the signal-to-noise ratio is, the larger
the percentage of explained variance is. Even when the signal-to-noise
ratio is small, for example, 0.18, PTA still has good estimation of
informative features which are close to the true values.

%
%
\begin{table}[b]
\tabcolsep=0pt
\caption{Design 4: studying the performance of PTA with the presence of
``noisy subjects''. Simulations are performed according to (\protect\ref
{simequ1})} 
\label{design4} 
{\fontsize{8.5pt}{11pt}\selectfont{
\begin{tabular*}{\tablewidth}{@{\extracolsep{\fill}}l cccccccc@{}} 
\hline
\textbf{Case} & $\bolds{P}$ & $\bolds{\mathcal{P}}$ & $\bolds{N}$ &$\bolds{T}$
& $\bolds{\mathcal{N}}$ & $\bolds{\mathcal{N}/ N}$
& \textbf{Nonzero features} & \textbf{Explained variance}\\ 
\hline
1 & 400 & 300 & 50 & 50 & $5$ & 0.1 & (150.0, 100.3, 50.0) $\pm$ (0.00,
0.95, 0.00) & 0.869 $\pm$ 0.0001\\ 
2 & 400 & 300 & 50 & 50 & $10$ & 0.2 &(150.0, 100.5, 50.0) $\pm$ (0.00,
1.08, 0.00) & 0.768 $\pm$ 0.0001\\
3 & 400 & 300 & 50 & 50 & $15$ & 0.3 &(150.0, 100.0, 50.0) $\pm$ (0.00,
0.00, 0.00) & 0.668 $\pm$ 0.0002\\
4 & 400 & 300 & 50 & 50 & $20$ & 0.4 &(150.0, 100.5, 50.0) $\pm$ (0.00,
1.58, 0.00) & 0.568 $\pm$ 0.0002\\
\hline
\end{tabular*}}}
\end{table}

Fourth, we study the effect of ``noisy subjects'' on the performance of
the method. In real data sets, some subjects can be ``outliers.'' It may
be due to the existence of unknown subpopulations or experimental
errors. We design simulations with the presence of ``noisy subjects'' as
illustrated in Table~\ref{design4}. We fix the total number of subjects
$N$ as 50, and increase the number of ``noisy subjects'', which is
denoted by $\mathcal{N}$. In this design, we assume the error term in
(\ref{simequ1}) follows a normal distribution, which is $\epsilon_{\mathit{npt}}
\sim N(0, 0.1)$. We repeat the simulation 10 times. One can see that
from Table~\ref{design4}, the ability of detecting informative
features is robust against the percentage of ``noisy subjects'', though
the percentage of explained variance decreases when the percentage of
``noisy subjects'' increases.

Fifth, we study the performance of PTA in terms of less number of time
points and larger number of features. Once a biological sample is
collected, on one hand, the lab technician may label it at once so that
thousands of genes can be scanned simultaneously. While on the other
hand, it is very time and labor intensive to extract samples at many
time points. Thus, we want to investigate the performance of PTA where
there are a larger number of features and a smaller number of time
points. We design the simulation as shown in Table~\ref{design5}. We
reduce the number of time points to 20 and 10, respectively, and
increase the number of features to 1000 and 10,000, respectively. We
simulate data based on
(\ref{simequ1}) with $\epsilon_{\mathit{npt}} \sim N(0, 0.1)$. We repeat the
simulation 10 times. The average performance of PTA under each setting
is shown in Table~\ref{design5}. One can see even with a smaller number
of time points and a larger number of features, PTA can still do a good
job. Nevertheless, increasing the number of time points and reducing
the number of noisy features help the performance of PTA.

%
%
\begin{table}
\tabcolsep=0pt
\caption{Design 5: studying the performance of PTA in terms of a
smaller number of time points and a larger number of features.
Simulations are performed according to (\protect\ref{simequ1})}
\label{design5} 
{\fontsize{8.5pt}{11pt}\selectfont{
\begin{tabular*}{\tablewidth}{@{\extracolsep{\fill}}l ccccc@{}} 
\hline
\textbf{Case} & $\bolds{P}$ & $\bolds{N}$ &$\bolds{T}$
& \textbf{Nonzero features} &\textbf{Explained variance}\\ 
\hline
1 & \hphantom{00,}400 & 1 & 20 &$(150.9, 100.3, 50.5) \pm(1.37, 0.95, 1.18)$& 0.810
$\pm$ 0.0008\\ 
2 & \hphantom{00,}400 & 1 & 10 &$(150.0, 101.6, 50.0) \pm(3.27, 3.20, 3.16)$& 0.730
$\pm$ 0.0026\\
3 & \hphantom{0,}1000 & 1& 20 &$(151.5, 101.0, 50.1) \pm(1.51, 1.41, 0.32)$& 0.776
$\pm$ 0.0007\\
4 & \hphantom{0,}1000 & 1& 10 &$(148.5, 96.9, 49.1) \pm(2.84, 2.81, 3.81)$& 0.688
$\pm$ 0.0019\\
5 & 10,000 & 1& 20 &$(151.7, 100.0, 49.7) \pm(1.70, 0.00, 1.70)$& 0.473
$\pm$ 0.0006\\
6 & 10,000 & 1 & 10 &$(148.0, 98.6, 47.6) \pm(3.20, 3.27, 3.37)$& 0.370
$\pm$ 0.0017\\
\hline
\end{tabular*}}}
\end{table}

\begin{figure}

\includegraphics{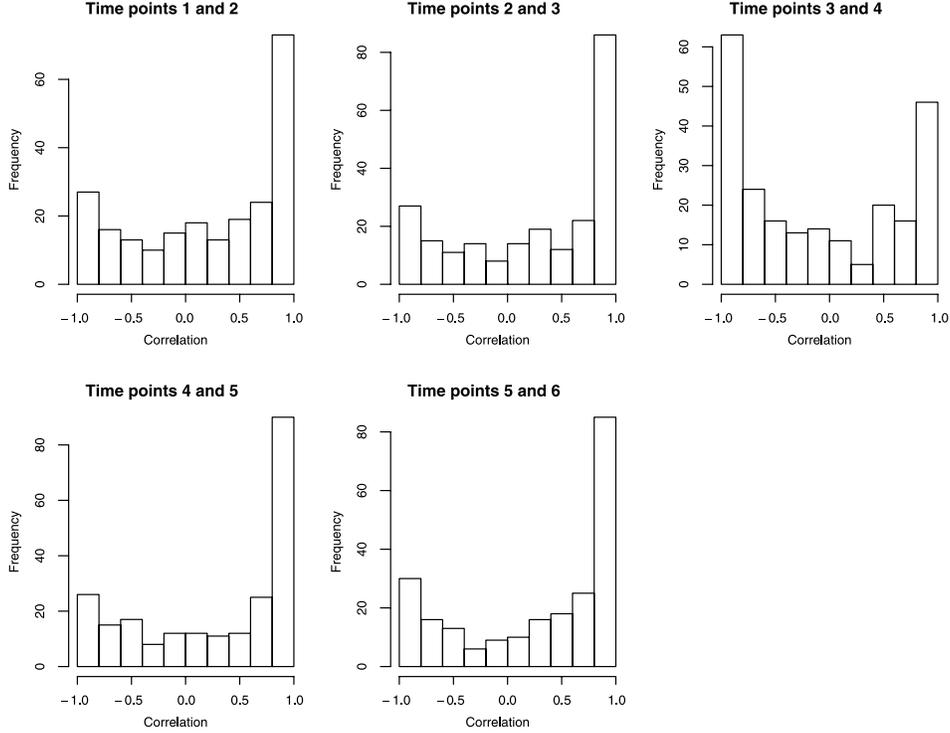}

\caption{Time-series correlation in noise in the circadian rhythm data
set in Section \protect\ref{leaf}. Histograms of values for correlation between
adjacent time points for 288 genes.} \label{noisecircadian}
\end{figure}

We have been assuming the error term $\epsilon_{\mathit{npt}}$ follows a normal
distribution so far. However, we want to know how much the method is
affected if the noise term $\epsilon_{\mathit{npt}}$ has some serial correlation
within observations from the same subject. The circadian rhythm data
set that we consider in Section~\ref{leaf} shows evidence of such
correlation. As shown in Figure~\ref{noisecircadian}, many genes have
high time-series correlation. We use the model of
(\ref{simequ1}) but generate $\epsilon_{\mathit{npt}}$ from the first order
auto-regression model with correlation $\rho=\operatorname
{corr}(\epsilon_{\mathit{npt}}, \epsilon_{np(t+1)})$, denoted by $\operatorname
{AR}(1, \rho)$. The design is illustrated in Table~\ref{design5}. We
study the performance of PTA with respect to different levels of
``noisy features'', different numbers of subjects and different
correlation levels in the error term. The results are shown in
Table~\ref{design6}. One can see that PTA has good performance in terms
of serial correlation. PTA is robust against the percentage of
noninformative features with the number of subjects, the number of time
points and the error term held fixed. PTA achieves a good performance
even with a small number of subjects under the condition that the noise
term has time-series correlation. Overall, the performance of PTA is
robust against serial correlation. Nevertheless, PTA has better
performance when the noise term has smaller serial correlation.

%
%

\begin{figure}

\includegraphics{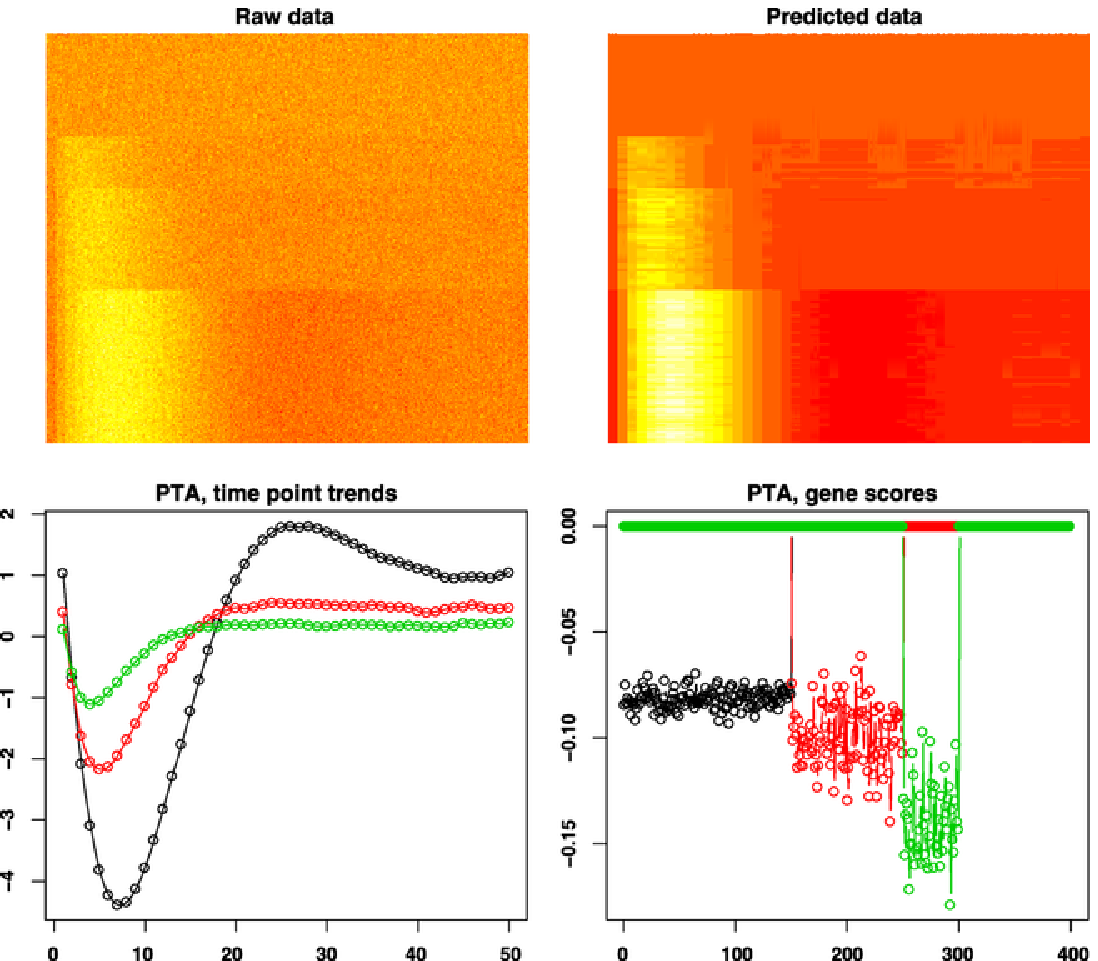}

\caption{One example in design 6. PTA on the simulated data with
multiple correlated patterns and noisy genes. Data was simulated
according to (\protect\ref{simequ2}) with $\epsilon_{\mathit{npt}} \sim N(0,
0.1)$, $P=400$, $N=10$ and $T=50$. The top-left panel shows the
simulated raw data, with rows indicating genes, columns indicating
samples ordered by time points and samples from the same time point
grouped together. The top-right panel shows the predicted data with
three PTs. The bottom-left panel shows the identified top three PTs by
PTA: black, the first PT which extracts the dominant pattern
$\exp(-t)\cdot\sin(0.5\cdot\pi\cdot t )$; red, the second PT which
extracts the second dominant pattern $\exp(-2 t)\cdot\sin(0.5\cdot\pi
\cdot t )$; green, the third PT which extracts the third dominant
pattern $\exp(-3 t)\cdot\sin(0.5\cdot\pi\cdot t )$. The bottom-right
panel shows scores of genes for their contributions to the top three
PTs: black, the first PT; red, the second PT; green, the third PT.
Tuning parameters are obtained by tenfold cross-validation.}
\label{figure2}
\end{figure}

The three components of the time signal are orthogonal functions in
(\ref{simequ1}). We want to know what happens if the latent time
patterns are ``correlated''. For instance, the expression of some genes
may decay or increase with time but with different rates and perhaps
oscillatory behavior. The principal trends from the burn data set that
we consider in Section~\ref{burn} are an example of such correlations among
latent time course patterns. We design the following experiment. Let
$\mathbf{Y}$ indicate longitudinal high-throughput gene expression from
$P$ genes, $N$ subjects and $T$ time points. We assign the values of
elements in $\mathbf{Y}$ as follows:
%
%
\begin{equation}\label{simequ2}
y_{\mathit{npt}} =\sin(0.5\cdot\pi\cdot t ) \cdot\sum
_{k=0}^{2} w_{k,
p}\cdot\exp(
\xi_k t) + \epsilon_{\mathit{npt}},
\end{equation}
where $\xi_0 <0$, $\xi_1<0$, $\xi_2 <0$, $p$ is the gene, $n$ is the
subject, $t$ is time, $\epsilon_{\mathit{npt}}$ is the error term, $w_{0,p}$ is
$I(0<p\le150)$, $w_{1,p}$ is $I(150<p\le250)$, $w_{2,p}$ is
$I(250<p\le300)$ and $I$ denotes the indicator function. We set $\xi_0
= -1$, $\xi_1 = -2$, $\xi_2 = -3$ in the simulation. We assume that the
error term follows a normal distribution $N(0, 0.1)$ as shown in
Table~\ref{design6}. We repeat the simulation 10 times. One can see
that PTA
%
%
\begin{sidewaystable}
\textwidth=\textheight
\tablewidth=\textwidth
\caption{Design 6: studying the performance of PTA when the noise term
has time series correlation. Simulations are performed according to~(\protect\ref{simequ1})}
\label{design6} 
\begin{tabular*}{\tablewidth}{@{\extracolsep{\fill}}l d{3.0}cd{2.0} ccc cc@{}} 
\hline
\textbf{Case} & \multicolumn{1}{c}{$\bolds{P}$} & $\bolds{\mathcal{P}}$
& \multicolumn{1}{c}{$\bolds{N}$} &$\bolds{T}$
& $\bolds{\mathcal{P}/{P}}$ & $\bolds{\epsilon
_{\mathit{npt}}}$ & \textbf{Nonzero features} &
\textbf{Explained variance}\\ 
\hline
1 & 400 & 100 & 1 & 50 & 0.25 & $\operatorname{AR}(1)$, $\rho=0.8$ &
(150.4, 101.2, 51.0) $\pm$ (0.97, 1.40, 1.70) &0.812 $\pm$ 0.0017\\ %
2 & 500 & 200 &1 & 50 & 0.40 & $\operatorname{AR}(1)$, $\rho=0.8$ &
(150.2, 100.0, 50.0) $\pm$ (0.42, 0.00, 0.00) &0.800 $\pm$ 0.0024\\
3 & 600 & 300 & 1 & 50 & 0.50 & $\operatorname{AR}(1)$, $\rho=0.8$ &
(150.3, 100.5, 50.1) $\pm$ (0.67, 0.85, 0.32) &0.788 $\pm$ 0.0017\\
4 & 1000 & 700 & 1 & 50 & 0.70 & $\operatorname{AR}(1)$, $\rho=0.8$ &
(150.6, 101.6, 51.5) $\pm$ (1.07, 1.96, 2.12) &0.745 $\pm$ 0.0014\\
5 & 400 & 100& 5 & 50 & 0.25 & $\operatorname{AR}(1)$, $\rho=0.8$ &
(150.0, 100.4, 50.7) $\pm$ (0.00, 0.97, 1.49) &0.900 $\pm$ 0.0007\\
6 & 400 & 100 & 10 & 50 & 0.25 & $\operatorname{AR}(1)$, $\rho=0.8$ &
(150.4, 100.1, 50.0) $\pm$ (1.26, 0.32, 0.00) &0.914 $\pm$ 0.0004\\
7 & 400 & 100 & 40 & 50 & 0.25 & $\operatorname{AR}(1)$, $\rho=0.8$
&(150.0, 100.7, 50.1) $\pm$ (0.00, 1.64, 0.32) &0.925 $\pm$ 0.0001\\
8 & 400 & 100 & 1 & 50 & 0.25 & $\operatorname{AR}(1)$, $\rho=0.1$ &
(150.0, 100.8, 50.2) $\pm$ (0.00, 1.23, 0.63) &0.838 $\pm$ 0.0003\\ %
9 & 400 & 100& 1 & 50 & 0.25 & $\operatorname{AR}(1)$, $\rho=0.5$ &
(150.1, 100.8, 50.6) $\pm$ (0.32, 1.03, 1.26) &0.833 $\pm$ 0.0003\\
\hline
\end{tabular*}
\end{sidewaystable}
works when underlying patterns are correlated. We plot one example
which is shown in Figure~\ref{figure2}. This example shows that PTA can
also detect the true informative features and extract the true
patterns. Besides, increasing the number of samples helps the
performance of PTA as shown in Table~\ref{design7}. Comparing Table~\ref{design7} with Tables~\ref{design1} and \ref{design2}, one can see
that PTA achieves better performance when the underlying time-course
patterns are less correlated.

%
%

%
%
\begin{table}
\tabcolsep=0pt
\def\arraystretch{0.9}
\caption{Design 7: studying the performance of PTA when the underlying
time-course patterns are correlated. Simulations are performed
according to (\protect\ref{simequ2})} 
\label{design7} 
{\fontsize{8.5pt}{11pt}\selectfont{
\begin{tabular*}{\tablewidth}{@{\extracolsep{\fill}}l d{4.0}cd{2.0}cccc@{}} 
\hline
\textbf{Case} & \multicolumn{1}{c}{$\bolds{P}$} & $\bolds{\mathcal{P}}$
& \multicolumn{1}{c}{$\bolds{N}$} &$\bolds{T}$
& $\bolds{\epsilon_{\mathit{npt}}}$ & \textbf{Nonzero
features} &\textbf{Explained variance}\\ 
\hline
1 & 400 & 100 & 1 & 50 & $N(0, 0.1)$ & (153.0, 97.9, 45.2) $\pm$ (2.21,
3.28, 0.63) & 0.530 $\pm$ 0.0088\\
2 & 500 & 200 & 1 & 50 & $N(0, 0.1)$ & (151.1, 96.9, 45.0) $\pm$ (3.38,
2.88, 0.00) & 0.477 $\pm$ 0.0125\\
3 & 600 & 300 & 1 & 50 & $N(0, 0.1)$ & (151.9, 97.0, 46.9) $\pm$ (3.41,
2.62, 3.25) & 0.432 $\pm$ 0.0086\\
4 & 1000 & 700 & 1 & 50 & $N(0, 0.1)$ & (151.7, 96.9, 45.2) $\pm$
(2.45, 3.07, 0.63) & 0.315 $\pm$ 0.0078\\
5 & 400 & 100 & 5 & 50 & $N(0, 0.1)$ & (152.3, 102.1, 50.2) $\pm$
(2.11, 2.13, 3.26) & 0.543 $\pm$ 0.0042\\
6 & 400 & 100 & 10 & 50 & $N(0, 0.1)$ & (150.8, 103.0, 50.0) $\pm$
(1.48, 1.56, 0.82) & 0.550 $\pm$ 0.0013\\
7 & 400 & 100 & 40 & 50 & $N(0, 0.1)$ & (150.1, 100.7, 50.0) $\pm$
(0.32, 1.64, 0.00) &0.551 $\pm$ 0.0007\\ 
\hline
\end{tabular*}}}
\end{table}

\begin{figure}[b]

\includegraphics{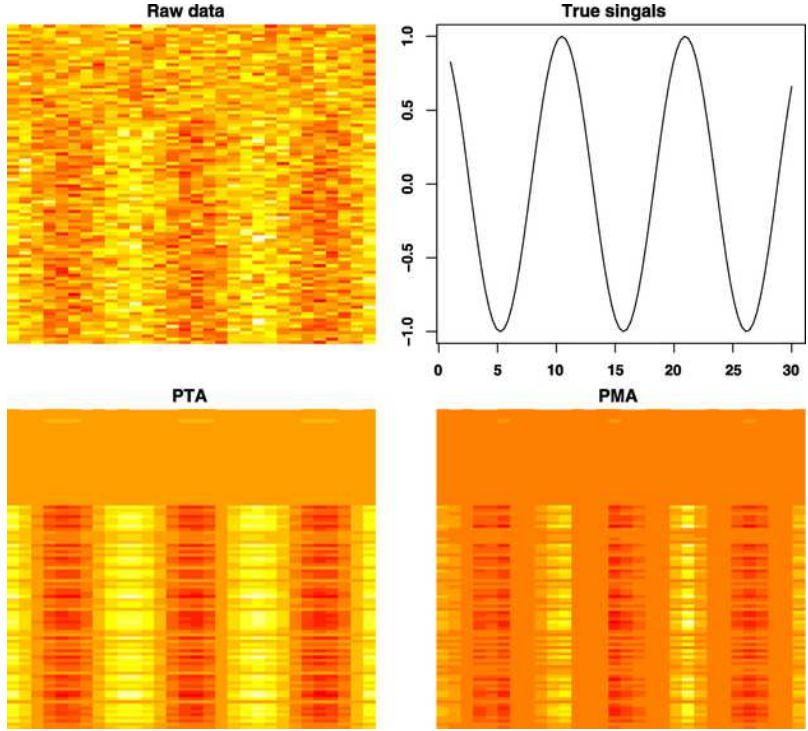}

\caption{Comparison of PTA and PMA. Data is simulated according to
procedure in (\protect\ref{simequ3}). The top-left panel shows the
simulated raw data, with rows indicating genes and columns indicating
samples ordered by time points. The top-right panel shows the true
signal. The bottom-left panel shows the predicted data using PTA. The
bottom-right panel shows the predicted data using PMA.} \label{figure3}
\end{figure}

Finally, we show why we need the penalty for smoothing in the model
(\ref{matrixop}) instead of using the Penalized Matrix Analysis [PMA;
cf. \citet{witten2009penalized}]. To compare the performance of
PTA and
PMA in extracting time-course patterns of gene expression with noisy
genes, we simulate a gene expression data set consisting of genes with
time-course patterns and noisy genes. Let $\mathbf{Y} = (y_{pt})$ be a
gene expression matrix with 100 genes and 30 time points. Elements of
matrix $Y$ are simulated according to the following model:
\begin{equation}
\label{simequ3}
y_{pt} = \cases{\cos(0.6 t) + \varepsilon_{pt},
& \quad$p \le70$,
\cr
\varepsilon_{pt}, &\quad$p>71$,}
\end{equation}
where $p$ is the indicator of gene, $t$ is time and $\varepsilon_{pt}
\sim N(0, 1)$. The heatmap of simulated raw gene expression is plotted
in the top-left panel of Figure~\ref{figure3}. Rows show genes and
columns indicate time points. The simulated data are represented in the
top-left panel of Figure~\ref{figure3}. The true time-course pattern is
%
%
illustrated in the top-right panel of Figure~\ref{figure3}, which is
the $\cos(0.6t)$ curve. We applied PTA and PMA on this simulated data
using single component decomposition. The prediction of PTA is obtained
by $\hat{\mathbf{a}}\hat{\bolds{\theta}}\mathbf{B}^\mathsf{T}$, and
presented by a heatmap in the bottom-left panel of Figure~\ref{figure3}. The prediction of PMA is obtained by $\hat{d}\hat{\mathbf
{u}}\hat{\mathbf{v}}^\mathsf{T}$, where $\{\hat{\mathbf{u}},\hat{\mathbf
{v}}\}=\arg\max_{\mathbf{u},\mathbf{v}}\mathbf{u}^\mathsf{T}\mathbf
{Y}\mathbf{v}$, subject to $\|\mathbf{u}\|_1\le c_1$, $\|\mathbf{u}\|
^2_2 =1$, $\|\mathbf{v}\|_1\le c_2$, $\|\mathbf{v}\|_2^2 =1$, $\hat
{d}=\hat{\mathbf{u}}^\mathsf{T}\mathbf{Y}\hat{\mathbf{v}}$. The
%
%
\begin{figure}

\includegraphics{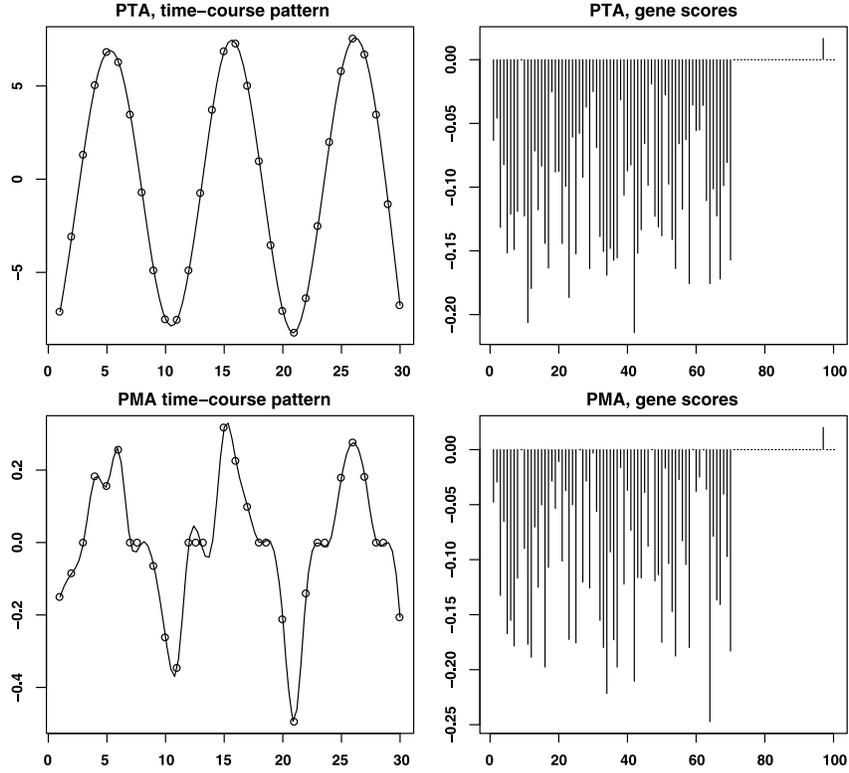}

\caption{Comparison of time-course patterns and scores identified by
PTA and PMA on the simulated data based on (\protect\ref{simequ3}). The
top-left panel shows the principal trend identified by PTA. Top-right
panel shows the gene scores of PTA. The bottom-left panel shows the
time-course pattern identified by PMA. The bottom-right panel shows the
gene scores of PMA.} \label{figure4}
\end{figure}
prediction of PMA is presented by a heatmap in the bottom-right panel
of Figure~\ref{figure3}. One can see that the prediction of PTA
extracts a clear cosine pattern, while the prediction of PMA is a
segmented curve. We also plot the time-course patterns and scores of
genes which are identified by PTA and PMA, respectively, in Figure~\ref{figure4}. Overall, the results suggest that PTA has better performance
than PMA in extracting time-course patterns of gene expressions.

\section{Circadian rhythm gene expression data}\label{leaf}
Circadian rhythms are biological processes that display endogenous and
entrainable oscillation of about 24 hours. These rhythms are driven by
a circadian clock. Circadian rhythms have been widely observed in live
organisms, including plants, animals, fungi and cyanobacteria.
Circadian regulations play important roles to maintain normal living of
live organisms. In particular, disruption to rhythms in humans may
result in a number of disorders, for example, bipolar disorder and
sleep disorders, and in the longer term is believed to have significant
adverse health consequences on peripheral organs outside the brain,
particularly in the development or exacerbation of cardiovascular
disease. Thus, circadian regulatory systems have attracted a lot of
attention for research and provide a good example to validate our PTA
approach. Here, to validate PTA, we use a circadian rhythm gene
expression data set from a well-studied pattern organism---arabidopsis
thaliana rosette.

\citet{blasing2005sugars} generated time-course gene expression data
from arabidopsis thaliana rosettes in a light-dark treatment
experiment. ATH1 arrays (Affymetrix Arabidopsis ATH1 Genome Array,
\href{http://www.affymetrix.com}{www.affymetrix.com}) were used to
measure gene expression to study how
diurnal cycle affects gene expression of arabidopsis thaliana rosettes.
Arabidopsis thaliana rosettes were harvested six times during
12-h-light/12-h-dark treatments. Three replicate samples were collected
4, 8 and 12 hours into the light period and 4, 8 and 12 hours into the
night. The time points were presented by \mbox{4, 8, 12, 16, 20, 24 h}. Gene
expression data were preprocessed by MAS [MicroArray Suite Software;
\citet{hubbell2002robust}] to evaluate probe set signals of the array.
The generated data files were further processed by RMA (Robust
Multi-array Average, from R-package Affy, \href{http://www.bioconductor.org}{http://www.bioconductor.org})
to normalize and estimate signal intensities. Gene expression values
were centered with mean 0. We preselected 228 genes with known
circadian regulation functions. This data set contains 228 genes, 6
time points and 3 replicates. Thus, to analyze circadian gene
expression pattern according to time, the standard PCA cannot be
applied while PTA should be used. We solve the following optimization
problem to extract the underlying time-course gene expression patterns
by PTA:
\[
\min_{\mathbf{A}, \bolds{\Theta}} L(\mathbf{A}, \bolds{\Theta}|\mathbf
{Y})\quad\mbox{such that}\quad\bolds{\Theta} \bolds{\Omega} \bolds
{\Theta}^\mathsf
{T}\le
c_1 \quad\mbox{and}\quad\|\mathbf{A}\|_2 =1.
\]

\begin{figure}

\includegraphics{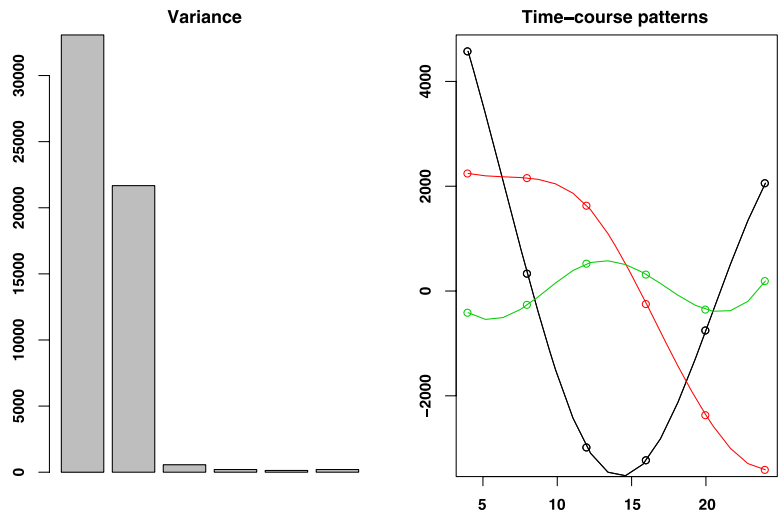}

\caption{PTA for the gene expression of genes with known circadian
regulated functions in Arabidopsis thaliana rosette. The left panel
shows variances for each component. Each bar shows one component,
ordered by the rank of component. The height of each bar shows the
variance for each component. The right panel shows the top three PTs.
Black indicates the first PT. Red indicates the second PT. Green
indicates the third PT.} \label{figure5}
\end{figure}

We plot the variance of each component in the left panel of Figure~\ref{figure5}. One can see that the first two PTs occupy much larger
variance than the rest of components. We also plot the first three PTs
in the right panel of Figure~\ref{figure5}. From the picture, one can
see that the first two PTs extract the patterns with greatest
longitudinal variations, which is consistent with the results in the
left panel of Figure~\ref{figure5}. In particular, the first cosine
shape pattern is the dominant shape, which agrees with the time pattern
of light in the experiment. For the first PT, 96 of the total 228 genes
%
%
have positive weights, while 132 of them have negative weights. Those
genes with positive weights share the same up and down directions with
the principal trend, while those with negative weights share the
opposite up and down directions. For example, the gene CAT2 with a
positive weight has a peak at the fourth hour. An independent
experiment has revealed the CAT2 mRNA accumulated to a peak four hours
after the onset of illumination and then declined, when etiolated
seedlings were illuminated [\citet{zhong1994interactions}]. Our
principal trend analysis suggests that light can be depicted as an
activator of CAT2 gene expression, probably through the action of the
phytochrome sensory system. The circadian clock could be envisaged as a
permissive regulator with respect to light, allowing induction of CAT2
at dawn.

\section{Gene expression of a burn cohort}\label{burn}
Despite ongoing improvements in resuscitation, care and outcomes, burn
injury remains a significant health and economic burden globally. The
current approach to the clinical management of these patients remains
limited by insufficient understanding of the pathobiology of the
disease. To describe the human genomic response to burns, a cohort of
burn patients were monitored and their gene expression was measured at
multiple time points (\href
{http://www.gluegrant.org}{www.gluegrant.org}). The longitudinal data
can be
divided into three stages---early stage (within one day to ten days
with three days median time), middle stage (eleven days to fourty-nine
days with nineteen days median time) and late stage (fifty days to more
than one year). Blood samples of burn patients were collected to
measure the gene expression by the Affymetrix HU133 Plus 2.0 arrays
[Zhang, Tibshirani and Davis
(\citeyear{Zhang2010survival,zhang2012classification})]. Each array
consisted of 54,675 probe sets. Gene expression data was normalized by
dChip [see \citet{Li2001}] and further reduced to 1000 probe sets with
the top 1000 highest coefficient of variation (CV, standard
deviation/mean). We take 28 surviving patients with
multiple-organ-failure scores (MOF) of three or less. Clinically, they
belong to the ``uncomplicated group.''

In this study, a preknown set of burn-related genes is not available.
An important task is to automatically identify the subset of genes
involved in burn response and extract their time-course expression
patterns. As suggested in the last simulation of Section~\ref{simulation}, not PMA in \citet{witten2009penalized}, but PTA is
suitable for dimension-reduction and feature selection in the
time-course gene expression data in a population of patients. Thus, we
solve the following optimization problem:
\[
\min_{\mathbf{A}, \bolds{\Theta}} L(\mathbf{A}, \bolds{\Theta}|\mathbf
{Y})\quad\mbox{such that}\quad\bolds{\Theta} \bolds{\Omega} \bolds
{\Theta}^\mathsf{T}
\le c_1,\qquad \|\mathbf{A}\|_1\le c_2\quad\mbox{and}\quad\|\mathbf{A}\|_2
= 1.
\]

\begin{figure}

\includegraphics{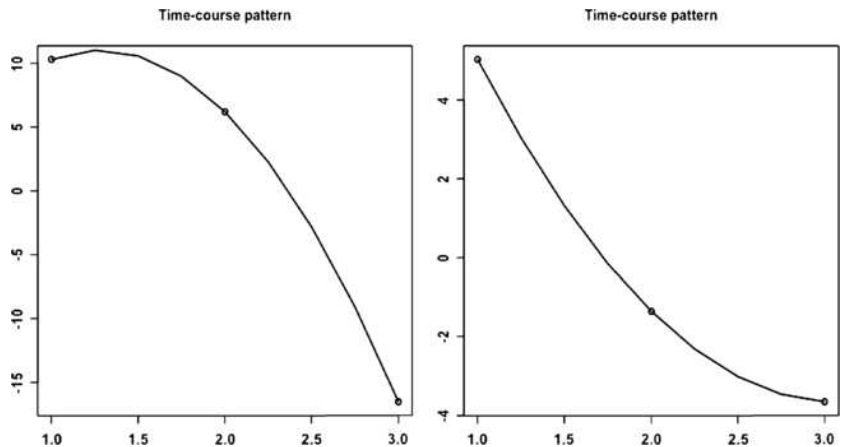}

\caption{PTA on burn data. Left: the first PT; right: the second PT.}
\label{figure6}
\end{figure}

\begin{figure}

\includegraphics{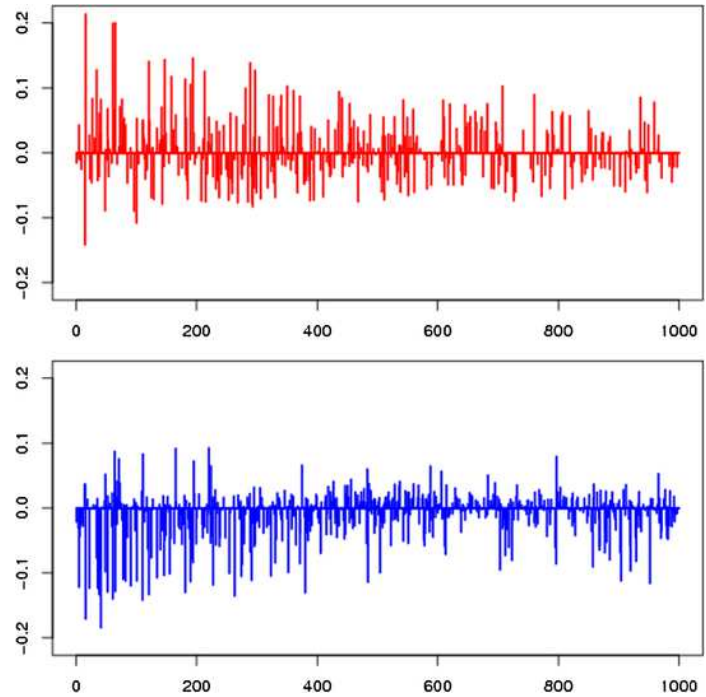}

\caption{PTA on burn data. The scores of genes for the first two PTs
are shown. The upper panel: the first PT; the lower panel: the second PT.}
\label{figure7}
\end{figure}

\begin{figure}[b]

\includegraphics{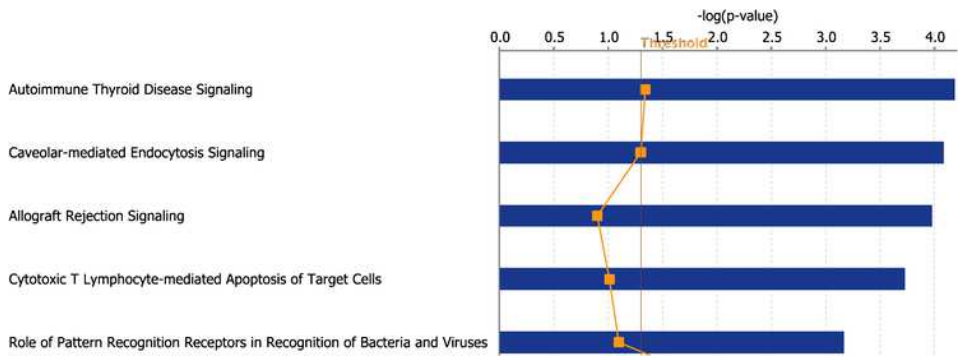}

\caption{PTA on burn data. Enriched canonical pathways for genes with
nonzero scores at the first PT.}
\label{figure8}
\end{figure}

\begin{figure}

\includegraphics{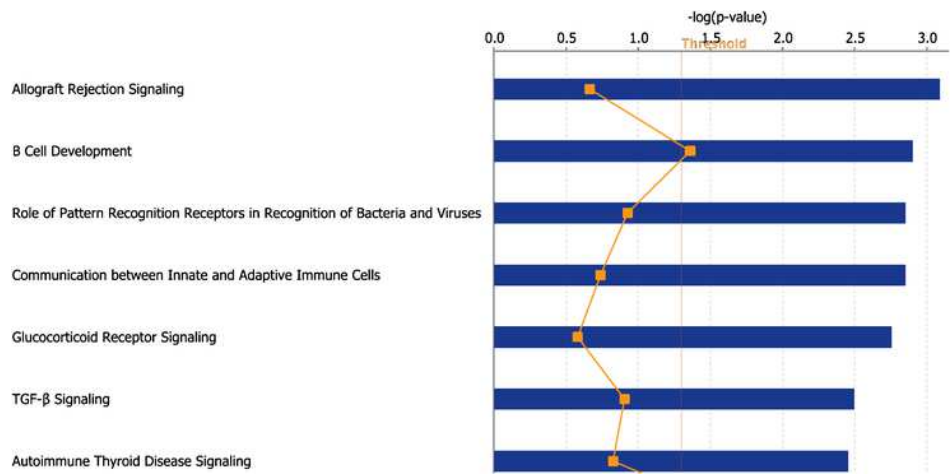}

\caption{PTA on burn data. Enriched canonical pathways for genes with
nonzero scores at the second PT.}
\label{figure9}
\end{figure}

After applying PTA on this data set of burn patients, we investigate
the extracted PTs and identified genes that have nonzero scores. The
top two PTs are shown in Figure~\ref{figure6}, and the scores of genes
%
%
are shown in Figure~\ref{figure7}. There are 400 genes with
contributions on the first PT, and 600 genes with contributions on the
second PT. We applied the enrichment analysis (Fisher's exact test)
%
%
using the Ingenuity Pathway Analysis (IPA) tool
(\url{http://www.ingenuity.com/}) on the genes with nonzero scores for the
first and second PTs, respectively. Top enriched canonical pathways are
shown in Figures~\ref{figure8} and \ref{figure9}. Genes in the
%
%
first PT have functions related to inflammatory responses, immune cell
trafficking, and cell-to-cell signaling and interaction functions.
Genes in the second PT have functions related to allograft rejection
signaling, B cell development, pattern recognition receptors in
recognition of bacteria and viruses, communication between innate and
adaptive immune cells, and cellular movement, growth and proliferation
functions. PTA provides an overview of the dynamics of genomic response
to burn injuries and extracts genes for further investigation to better
%
%
understand the pathobiology of burn disease. For example, gene IFIT1 is
one of the identified genes with nonzero scores in the first PT. It has
been shown that in mammalian cells, it is synthesized in response to
viral infection, and consequently assigns resistive activity against
viral invasion to cells [\citet{li2010interaction}]. It is common for
burn patients to suffer from infectious episodes, which are caused by
viruses [\citet{finnerty2006cytokine}]. Thus, PTA successfully
identified the key regulatory element IFIT1, which plays an important
role for the recovery of burn patients. Overall, PTA characterizes the
inflammatory transcriptome following a burn injury and identifies the
burn-induced immuno-inflammatory dysfunction and hyperinflammatory
response. Comprehensive understanding of the molecular mechanisms of
burn disease will ultimately lead to novel and profound advancements in
clinical care.

\section{Discussion}\label{discussion}
By virtue of matrix theory, PCA is an important methodology to study
data structure. It has been widely used to analyze clinical and
biological data. Combining the \emph{lasso} technique with matrix
decomposition, the regularized PCA method can be applied to
high-throughput genomic or proteomic data. In the case of a time-course
experimental scenario, such data sets have four dimensions (feature,
sample, time, gene/protein expression), traditional PCA or regularized
PCA cannot be used directly. We have developed a new principal trend
analysis (PTA) method for time-course data modeling. PTA incorporates
smoothing techniques for time and sparsity techniques for genes when it
is necessary. Our simulations and applications on real data sets show
that PTA works well and outperforms regularized PCA (PMA) on
time-course gene expression data.

The proposed PTA focuses on a population's time-course patterns.
However, it also works when the number of subjects equals one. If
subjects are too diverse and cannot be treated as one population, we
can apply PTA on each subject and identify personalized longitudinal
gene expression patterns and important genes.

In clinical and biological studies, people may collect different types
of data sets on the same patients or samples at multiple time points.
Those time-course data includes gene expression and protein abundance,
and clinical measurements of the same patients, etc. It will be
interesting to extract the associated relationships among features from
multiple longitudinal data sets. Developing methods to handle such
tasks will be an important future work.

\section*{Acknowledgments}

The authors thank Dr. Robert Tibshirani for many discussions and useful
comments. The authors also appreciate the Editor and referees for their
tremendous insightful suggestions and comments, as well as Dr.
Guilherme V. Rocha for his help on the proof of biconvex property in
Section~1 of the supplementary material [\citet{ZhaDav13}]. The authors wish to acknowledge the
efforts of many individuals at participating institutions of the Glue
Grant Program that generated the data reported here.

\begin{supplement}
\stitle{Supplement to ``Principal trend analysis for time-course data with applications
in genomic medicine''}
\slink[doi]{10.1214/13-AOAS659SUPP} 
\sdatatype{.pdf}
\sfilename{aoas659\_supp.pdf}
\sdescription{The supplementary material includes ``Proof of biconvex property'' and ``Derivation of PTA algorithm.''}
\end{supplement}

%

\printaddresses

\end{document}